\documentclass{article}

\usepackage{PRIMEarxiv}

\usepackage[utf8]{inputenc} 
\usepackage[T1]{fontenc}    
\usepackage{hyperref}       
\usepackage{url}            
\usepackage{booktabs}       
\usepackage{amsfonts}       
\usepackage{nicefrac}       
\usepackage{microtype}      
\usepackage{lipsum}
\usepackage{fancyhdr}       
\usepackage{graphicx}  
\usepackage{graphicx}
\usepackage{amssymb}
\usepackage{arydshln}
\usepackage{comment}
\usepackage{multirow}
\usepackage{colortbl}
\usepackage{breqn}
\usepackage{float}
\usepackage{amsthm}
\restylefloat{table}
\usepackage{placeins}

\usepackage{subfigure}
\usepackage{pifont}
\usepackage{enumitem}
\newcommand{\subscript}[2]{$#1 _ #2$}
\newcommand{\cmark}{\ding{51}}%
 \newcommand{\xmark}{\ding{55}}
\graphicspath{{media/}}     

\newtheoremstyle{sig}
  {}
  {}
  {\itshape}
  {}
  {\scshape}
  {.}
  {.5em}
  {#1 #2\thmnote{\quad(#3)}}

\theoremstyle{sig}

\newtheorem{remark}{Remark}

\title{CALEB: A Conditional Adversarial Learning Framework to Enhance Bot Detection

}

\author{
  George Dialektakis \\
  Aristotle University of Thessaloniki \\
  Department of Informatics \\
  Thessaloniki, Greece\\
  \texttt{gdialekt@csd.auth.gr} \\
   \And
  Ilias Dimitriadis \\
  Aristotle University of Thessaloniki \\
  Department of Informatics \\
  Thessaloniki, Greece\\
  \texttt{idimitriad@csd.auth.gr} \\
     \And
  Athena Vakali \\
  Aristotle University of Thessaloniki \\
  Department of Informatics \\
  Thessaloniki, Greece\\
  \texttt{avakali@csd.auth.gr} \\
}

\begin{document}
\maketitle

\begin{abstract}
The high growth of Online Social Networks (OSNs) over the last few years has allowed automated accounts, known as social bots, to gain ground. As highlighted by other researchers, most of these bots have malicious purposes and tend to mimic human behavior, posing high-level security threats on OSN platforms. Moreover, recent studies have shown that social bots evolve over time by reforming and reinventing unforeseen and sophisticated characteristics, making them capable of evading the current machine learning state-of-the-art bot detection systems. This work is motivated by the critical need to establish adaptive bot detection methods in order to proactively capture unseen evolved bots towards healthier OSNs interactions. In contrast with most earlier supervised ML approaches which are limited by the inability to effectively detect new types of bots, this paper proposes CALEB, a robust end-to-end proactive framework based on the Conditional Generative Adversarial Network (CGAN) and its extension, Auxiliary Classifier GAN (AC-GAN), to simulate bot evolution by creating realistic synthetic instances of different bot types. These simulated evolved bots augment existing bot datasets and therefore enhance the detection of emerging generations of bots before they even appear! Furthermore, we show that our augmentation approach overpasses other earlier augmentation techniques which fail at simulating evolving bots. Extensive experimentation on well established public bot datasets, show that our approach offers a performance boost of up to 10\% regarding the detection of new unseen bots. Finally, the use of the AC-GAN Discriminator as a bot detector, has outperformed former ML approaches, showcasing the efficiency of our end to end framework. 
\end{abstract}


\section{Introduction}
Over the last decade, Online Social Networks (OSNs) have prevailed as the default communication, interaction, and information sharing platforms. Recent analysis reveals that there are 4.65 billion social media users around the world as of April 2022, equating to almost than 70 percent of the eligible global population \cite{kepios_general}. 
A recent study \cite{kepios_twitter} has shown that Twitter, one of the most 'active' OSNs with more than 436.4 million monthly active users, has a +90\% growth of daily active users, confirming that OSNs have become an integral part of humans daily lives.
However, the openness and easily accessible OSNs interfaces has triggered the rise of automated accounts, also known as social bots \cite{ferrara2016rise}. Such accounts are machine or human operated software, either benign or malicious, that tend to imitate human behavior by posting content and interacting with humans in order to achieve their goals \cite{botomics2}. Since they are automated, they operate much faster than human users, carrying out useful functions without the need for human supervision.

Unfortunately, bots can also come in the form of malware, posing high-level security threats in OSN platforms \cite{botomics21}. Especially on Twitter, where humans are more open to expressing their opinions and views, social bots find an excellent ground to impact people's thoughts and mindsets. Spreading misinformation and fake news has become a major critical issue and it is widely recognized that public opinion is actively influenced in a meta-truth era at which online trust is at risk \cite{cgan5}. 
OSNs like Twitter, have already recognized the need to take measures against the social bots. In accordance with a study from The Washington Post from May 2018, Twitter has identified almost 10 million bot accounts \cite{wu2020using}. In addition, social bots are accountable for producing 35\% of the content posted on Twitter, as discovered by another study \cite{cgan8}. Even more recent studies have shown that during the COVID pandemic, most of the Twitter content related to this topic was posted by bots \cite{cmucovid, ferrara2020covid}. Thus, the high degree of social bots influence on Twitter and their ability to tamper with our social ecosystems massively, is an actual threat to our societies and democracies which struggle with high polarization and hateful speech in online debates \cite{botomics14,botomics15}. 

As social bots have increasingly shown malicious activity in OSNs, negatively affecting a large part of the global population, many earlier research studies have proposed techniques to detect bots \cite{cgan9}. These techniques mainly focused on supervised and unsupervised Machine Learning (ML) algorithms that aim at separating bot accounts from legitimate ones by several actions (eg. detecting fake content related to each account, examining the account profile itself, and investigating the network of the account) \cite{cresci2020decade}. 

Even though many research studies have made a significant effort to detect social bots, most of them suffer from a significant weakness induced by the evolutionary nature of social bots, as demonstrated by a recent research survey \cite{cresci2020decade}. Up to now, bot evolution and transformation remains a major obstacle in all bot detection methods. The evolution of bots also results in the appearance of new bot types, which adopt advanced features and cutting edge strategies that make their detection a struggling endeavour. Thus, as bots evolve and become smarter, they develop capabilities of hijacking and evading the state-of-the-art bot detection systems \cite{cresci2019better}. In addition, armies of malicious bots mimic human behavior and share more similar characteristics with legitimate human accounts \cite{decade5}, making them less distinguishable from actual humans. Finally, current state-of-the-art techniques follow a \textbf{reactive approach}, coming too late in advancing new bot detection solutions, only after evidence of new evolved bots \cite{bettersafe43}. In this context, research efforts always remain one step behind the malicious evolving bot development \cite{cresci2021coming}.

Recently, a new Adversarial Machine Learning (AML) approach for bot detection has gained ground, with techniques which aim to become \textbf{proactive} and capable of anticipating the evolution of bots and detecting their emerging generations \cite{cresci2021coming}. Such methods involve a specific class of ML Neural Networks, namely Generative Adversarial Networks (GANs) \cite{goodfellow2014generative}. GANs can be trained to produce artificial samples of evolved social bots representing and precautiously capturing next generations of malicious bots. 


This approach can offer insights about the vulnerabilities of existing bot detection systems, before even bot developers discover and effectively exploit their weaknesses \cite{cresci2016dna}. Despite the emergence of Adversarial techniques in bot detection, this approach is still at an early experimental stage and has not been fully explored yet. For instance, the existing AML approaches focus only on discriminating bots from humans and produce synthetic samples of general purpose bots ignoring their different types that exist in OSN platforms. However, recent research has shown that several kinds of bots have emerged with different intentions and objectives, featuring distinctive characteristics \cite{dimitriadis_botomics}. Therefore, it is of vital importance to construct a detection system capable of classifying evolved bots of multiple types.

Attempting to address the above open problems referring to the multi-type detection of bots and their evolution, this paper proposes CALEB, a Conditional Adversarial Learning Framework to proactively enhance bot detection by creating synthetic bot instances of multiple types that simulate evolved bots. These simulated evolved bots are then used to augment existing bot datasets and develop a robust detection system towards new unseen multi-type bots. In specific, the main contributions of this paper are as follows:
\begin{enumerate}[label=\subscript{\textbf{C}}{{\textbf{\arabic*}}})]
    \item \textbf{We propose a feature-based exploratory analysis of different bot types.}: This study takes a further step from co-authors previous work \cite{dimitriadis_botomics} and proceeds with an additional exploratory analysis of feature categories that validates and highlights the diversity between the different types of bots. \label{c1}
    \item  \textbf{We introduce a novel proactive methodology for the detection of multi-type evolving bots}: Targeting to capture the evolutionary nature of social bots, we propose the use of Conditional Generative Adversarial Networks (CGAN) and Auxiliary Classifier Generative Adversarial Networks (AC-GAN), which are trained to generate synthetic bot samples that simulate evolved social bots. These artificial bot instances are used to enhance advanced ML classifiers in order to make them robust against future generations of bots. To the best of the authors' knowledge, this approach has not been presented in literature before. \label{c2}
    \item \textbf{We implement an end-to-end Generative Adversarial bot detection framework}: To reduce the training overhead, we propose the use of an end-to-end Generative Adversarial model, namely AC-GAN, capable of executing two major tasks: the generation of synthetic bot samples of different bot types and the detection of existing and future generations of social bots. \label{c3}
    \item \textbf{We validate the efficiency of our multi-type bot detection framework on various public bot datasets}. The results of the experimental evaluation reveal that the proposed methodology achieves comparable performance to other state-of-the-art techniques on class imbalance scenarios. Additionally, extensive experiments demonstrate that our approach effectively addresses the issue of bot evolution, achieving up to 10\% performance boost in comparison to already established methods, that do not take into account the evolving nature of bots. \label{c4}
\end{enumerate}

The remainder of this paper is organized as follows: Section \ref{sec_related} outlines earlier related work. Section \ref{sec_exploration} presents an exploratory analysis based on Twitter bot datasets. Section \ref{sec_methodology} justifies the proposed methodology for the detection of multi-type evolving bots and Section \ref{sec_experiments} illustrates our experimental evaluation along with the core results. Finally, Section \ref{sec_conclusion} concludes this work and proposes some ideas for future research.

\section{Related Work} \label{sec_related}
Recent studies have shown that most existing approaches in the literature rely mainly on Machine Learning supervised methodologies \cite{davis2016botornot, lee2010uncovering, varol2017online, yang2019arming, yang2020scalable_8, yardi2010detecting}.
Supervised methods mainly refer to classification, where a model is trained to distinguish between social bot accounts and legitimate human ones. 
However, there are also some approaches that address the issue on a multi-class level where algorithms are trained to discriminate the different types of social bots \cite{botomics15, chu2012detecting, dimitriadis_botomics, cresci2020decade}. 
However, these mainstream methods are only effective on detecting the existing bots while failing to do so for their evolved versions. To address the challenge of bot evolution, where bots are adapted so that they can evade current detection systems, these early approaches develop improved systems only after evidence of new evolved bots have been observed, thus staying one step behind bot developers \cite{bettersafe43}.

Recently, a new family of ML algorithms has started to gain ground in the field of social bot detection, namely Adversarial Machine Learning (AML) \cite{cresci2021coming, kurakin2016adversarial}.
At their core, adversarial methods use Generative Adversarial Networks (GANs) to generate realistic synthetic bot representations that look like existing ones, augmenting the available datasets with advanced bot samples. For example, Bin Wu et al. (2019) \cite{wu2019detecting}, proposed a simple GAN to tackle the problem of imbalance between bot and legitimate samples presented in most public bot datasets. Specifically, a simple GAN was trained on a dataset \cite{decade5,varol2017online} composed of 2433 samples (only 18\% were bot accounts) to generate synthetic bot samples in order to enhance the original dataset. Then they trained a separate Neural Network on the augmented dataset to distinguish between bot and human accounts. The results showed that their GAN-based approach outperformed five state-of-the-art oversampling techniques. Another adversarial approach to solve the problem of class imbalance in bot data is presented in \cite{wu2020using}, where a Conditional Generative Adversarial Network (CGAN) is proposed that is able to control the specific class of samples being generated by feeding the GAN with auxiliary information about the class labels of the data \cite{mirza2014conditional}. This additional information comes from a Gaussian kernel density peak clustering algorithm which clusters the bot samples based on their features and assigns a different bot category per cluster. This process helps the CGAN create more realistic synthetic bot samples while eliminating the imbalances between and within bot class distributions. Experimental results revealed that the advanced CGAN outperformed state-of-the-art oversampling techniques such as Random Oversampling, SMOTE, and ADASYN, reaching an F1 score of 97.56\%. 
Despite the success of the above works in dealing with class imbalance, they only focus on generating artificial bot samples that are quite similar to the pre-existing ones, leaving their detection models vulnerable to advanced and evolved bots. Moreover, they do not consider identifying and generating different types of bots; instead, they only approach bot detection as a binary classification problem.

The first approach towards dealing with bot evolution is performed by Cresci et al. (2019) \cite{cresci2019better} where they propose a proactive adversarial bot detection method using genetic algorithms. In detail, they develop a novel genetic algorithm, namely GenBot, which has the ability to create evolved Twitter spambots whose behavior looks like the behavior of legitimate accounts. GenBot is then combined with a digital DNA behavioral modeling technique \cite{cresci2017social} and produces synthetic evolved generations of bots by focusing on the sequences of actions of spambots and legitimate accounts on Twitter. The key point of this work is the ability to produce adversarial samples, i.e., synthetic bot accounts, that, through thorough experimentation, prove capable of evading state-of-the-art bot detection systems such as those in \cite{cresci2017social,miller2014twitter}. This work reveals that the reactive schema that most detection algorithms follow makes them extremely vulnerable to future generations of bots. Thus, research has to move in the direction of proactive strategies which promote detection models to a priori adapt, develop more sophisticated techniques and become more robust to the upcoming generations of evolved bots.

Another work that attempts to address the issue of bot evolution is presented in \cite{jan2020throwing}, where STK Jan et al. (2020) propose a GAN-based framework with two Generators for producing advanced artificial bot samples and show that this approach requires only 1\% of labeled data to outperform existing methods. In this context, a distribution-aware data synthesis is proposed based on known legitimate accounts and limited bot ones. The main idea is to generate synthetic bot examples for the unoccupied regions in the feature space by differentiating "outlier regions" representing new bot variants and "clustered regions" representing legitimate users. For this reason, two Generators are used to create these two types of data in a different manner. The first Generator produces clustered samples by gradually decreasing the aggressiveness of the data synthesis as we get closer to the benign region. On the other hand, the other Generator is more aggressive to fill in the space. The Discriminator of the GAN is trained to distinguish real from artificial data and legitimate users from bots simultaneously. After training their GAN-based model on a real-world dataset containing network traffic data from Radware, they point out that the proposed model outperforms other state-of-the-art approaches such as an LSTM network and another GAN-based method, namely OCAN, needing only 1\% labeled data. However, as they state in their paper, the proposed model cannot capture bots embedded in the benign region and legitimate users that behave quite differently from most other users, leaving their detection system vulnerable to those types of accounts.

There are also some other works that use AML, such as in \cite{yin2018enhancing} where a GAN-based framework is proposed to enhance the performance of previous botnet detection methods by generating synthetic botnet samples and the work in \cite{ma2019detect} where a text-based GAN is developed to detect rumors and fake news. However, both approaches belong to different bot domains and do not focus on Twitter bots, which is the main purpose of this paper. 

Even though several attempts have been made using AML for Bot Detection, this field still remains considerably unexplored. For example, most Adversarial approaches \cite{yin2018enhancing,wu2019detecting,ma2019detect,wu2020using} propose a GAN as an oversampling method to increase the number of bot samples without considering their evolutionary nature and adaptation that make them capable of evading the current state-of-the-art detection systems. Other Adversarial methodologies \cite{cresci2019better,jan2020throwing} that attempt to deal with the evolving behavior of bots, only focus on distinguishing bots from legitimate accounts without taking into consideration the different existing types of bots. Last but not least, there are only three research studies in the literature \cite{yin2018enhancing,ma2019detect,jan2020throwing} that utilize the GAN's Discriminator as a bot detector, and they focus only on the binary classification problem. On the other hand, to detect bots, most GAN-based frameworks train additional ML classifiers to perform the final classification, adding additional training overhead in their pipeline.

\begin{table}[t]
\centering
\caption{Previous efforts on Adversarial Bot Detection. CALEB matches all specs, while competitors miss one or more of the features.}
\resizebox{0.7\columnwidth}{!}{

\begin{tabular}{c|c|c|c}
\textbf{Paper}                & \begin{tabular}[c]{@{}c@{}}\textbf{Multiclass }\\\textbf{Bot Detection}\end{tabular} & \textbf{Bot Evolution} & \begin{tabular}[c]{@{}c@{}}\textbf{Discriminator }\\\textbf{as Bot Detector}\end{tabular}  \\ 
\hline
C. Yin et al. (2018)          & \xmark                                                                & \xmark  & \cmark                                                                      \\ 
\hline
Bin Wu et al. (2019)          & \xmark                                                                & \xmark  & \xmark                                                                      \\ 
\hline
J Ma et al. (2019)            & \xmark                                                                & \xmark  & \cmark                                                                      \\ 
\hline
Cresci et al. (2019)          & \xmark                                                                & \cmark  & -                                                                                          \\ 
\hline
STK Jan et al. (2020)         & \xmark                                                                & \cmark  & \cmark                                                                      \\ 
\hline
Bin Wu et al. (2020)          & \xmark                                                                & \xmark  & \xmark                                                                      \\ 
\hline
\textbf{CALEB (our approach)} & \cmark                                                                & \cmark  & \cmark                                                                      \\
\hline
\end{tabular}
}
\label{Table_function}
\end{table}

In this work, we attempt to fill in the identified gaps, as can be seen in Table \ref{Table_function}, by proposing CALEB, a Conditional Adversarial Learning Framework to enhance bot detection. More specifically, a Conditional Generative Adversarial Network and an Auxiliary Classifier GAN is proposed to generate simulated evolved bot examples and help towards establishing a robust multi-class bot detection system capable of identifying the different existing types of social bots and their next generations. Moreover, we investigate whether synthetic data produced by GANs can boost the classification performance in multi-class imbalanced bot data. Finally, we evaluate the 
use of the Auxiliary Classifier GAN (AC-GAN), as an end-to-end bot detection system, that is able to both generate artificial samples of bots and detect their existing and evolved versions. To the best of our knowledge, this is the first approach using GANs towards multi-type bot detection.
\section{Data Exploratory Analysis} \label{sec_exploration}
It is well known that the performance of bot detection methods as well as generative models highly depends on the data used to train and evaluate the models. In this section, we discuss about the data we used in this work, as well as, we present an exploratory analysis regarding the features that describe the social accounts.
\subsection{Dataset} \label{sec_dataset}

\begin{table}[b]
\centering
\caption{Different bot types included in the utilized dataset along with the number of examples for each bot category.}
\label{bot_types_table}
\resizebox{0.6\columnwidth}{!}{%
\begin{tabular}{c|c|c} 
\hline
\textbf{Bot Class}                           & \textbf{Description}               & \textbf{Instances}  \\ 
\hline
\textit{Spam Bot}                            & Accounts that post spam content    & 17071               \\ 
\hline
\textit{Social Bot}                          & Bots that try to attract followers & 11653               \\ 
\hline
\textit{Political Bot}                       & Bots that deal with politics       & 497                 \\ 
\hline
\textit{Cyborg}                              & Human monitored bots               & 5891                \\ 
\hline
\textit{Self-declared}                       & Accounts that state they are bots  & 1198                \\ 
\hline
\textit{Other Bot}                           & Other type of simple bots          & 2109                \\ 
\hline
\textit{Human}                               & Genuine human accounts             & 30752               \\ 
\hline
\multicolumn{1}{l|}{\textbf{Total accounts}} & -                                  & \textbf{69171}      \\
\hline
\end{tabular}
}
\end{table}

 Taking into consideration the universality and diversity, in this work we decided to use a dataset that was already created in \cite{dimitriadis_botomics}. This dataset consists of 24 different datasets, which most of them are accessible under the Data Repository section of Botometer \cite{yang2019arming_5,yang2020scalable_8,davis2016botornot_42}, containing social accounts from Twitter, and two of them are the result of a manual account search from Twitter. In the beginning, the original datasets were identified only by a binary label, human or bot. However, since the goal of this paper is to generate synthetic bot samples of multiple types, we followed the same bot type categorization as in \cite{dimitriadis_botomics}, where bots are divided into six categories: spam bots, social bots, political bots, cyborgs, self-declared bots, and other bots. In addition, we also have the human accounts. The different types of bots along with the number of examples for each social account type are shown in Table \ref{bot_types_table}.

\noindent Each social account in the dataset is represented by 310 features from five different categories, which is the result of a feature extraction process performed in \cite{dimitriadis_botomics}. Specifically, there are 182 Content features, 58 Sentiment features, 29 Temporal features, 28 User features, and 13 Hashtag Correlation Features.

\subsection{Feature Distribution Exploratory Analysis}
\begin{figure}[h!]
     \centering
     \subfigure[Temporal Features]{\label{temporal_pdf}\includegraphics[scale=0.33]{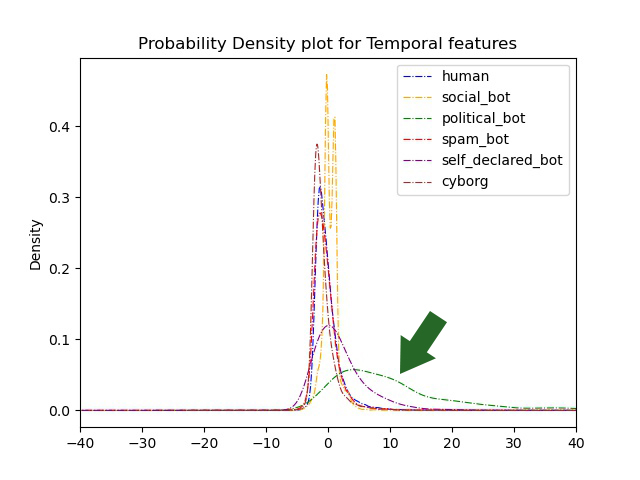}}
     \subfigure[User Features]{\label{user_pdf}\includegraphics[scale=0.33]{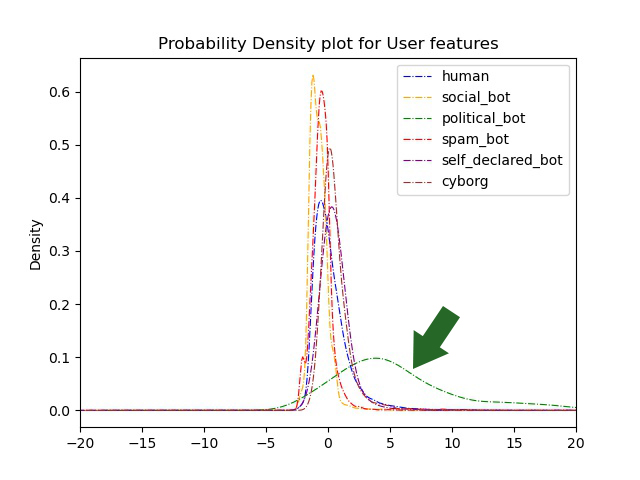}}
      \subfigure[Content Features]{\label{content_pdf}\includegraphics[scale=0.33]{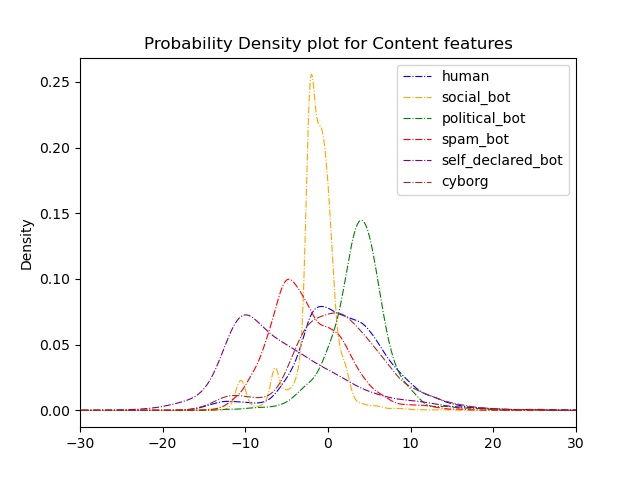}}

    \caption{Example of Probability Distribution Plots per Feature Category for different bot types.}
    \label{pdf_plots}
\end{figure}

As mentioned above, the dataset we used contains information about seven distinct account types. In this work, we decided to discard the other bots category, since it may contains bots from various bot types and therefore makes the analysis of the data more complicated. At this point, we wanted to examine how the features of each bot type differ compared to the others and how this distinction benefits the bot categorization we have considered. Since each feature category in our data consists of numerous features, it is infeasible to create plots to visualize their distribution. For that reason, we use Principal Component Analysis (PCA) \cite{pca}, to reduce the dimensions of our dataset from 310 (which is the original number of features) to 5, one dimension for each feature category. In other words, each feature category is projected into one dimension. Therefore, each social account in our dataset now consists of five features. After applying dimensionality reduction, we construct Probability Distribution Function (PDF) plots for each feature category. Our goal is to examine and compare the distribution of each feature category between the different types of bots, as stated in \textbf{(}\ref{c1} paper contribution. To this end and due to the lack of space, we present an example of the distribution for the temporal, content and user features, as illustrated in Figure \ref{pdf_plots}. 

\begin{remark}
\textbf{
The temporal and user features of political bots follow a distribution that significantly deviates from the distribution of the other bot types, as illustrated in Figure \ref{temporal_pdf} and \ref{user_pdf}, respectively.}
\end{remark}

\noindent In Figure \ref{temporal_pdf} and \ref{user_pdf}, we observe that the distribution of temporal and user features for political bots, respectively, differs from the other bots as it is centered around 5, while the curves for the other bot types are centered around zero, and has six times less height compared to the other curves. In terms of temporal features, this significant difference is somewhat expected as political bots diverge from the other classes since they are commonly more active during political events such as elections, etc, while the other bot types can be active on a daily basis.

\begin{remark}
\textbf{
The biggest divergence between the distribution curves is presented in the content features, as shown in Figures \ref{content_pdf}.
}
\end{remark}

\noindent Figure \ref{content_pdf} shows an example where the PDF distribution of the content features highly differs between the different bot types. It is obvious that there is no distribution that is identical to another. On the other hand, for instance, in the user and temporal features, most bot types follow a similar distribution. This consideration leads us to our final key observation.

\begin{remark}
\textbf{
Content features is probably the most important feature category among the five we have considered for a classifier to discriminate between the different classes of bots.
}
\end{remark}

\noindent As mentioned above, the PDF curves of the content features present the biggest difference between the different classes of social bots. This leads to the conclusion that this type of features contain more important information than the other categories, thus having more power to discriminate the different bot classes than features belonging to the other categories.
\section{Methodology} \label{sec_methodology}
In this section, we thoroughly describe our proposed methodology to address the unresolved issue of multi-type bot evolution.
    
\subsection{Overall Process}
\begin{figure}[t]

    \centering
    \includegraphics[scale=0.3]{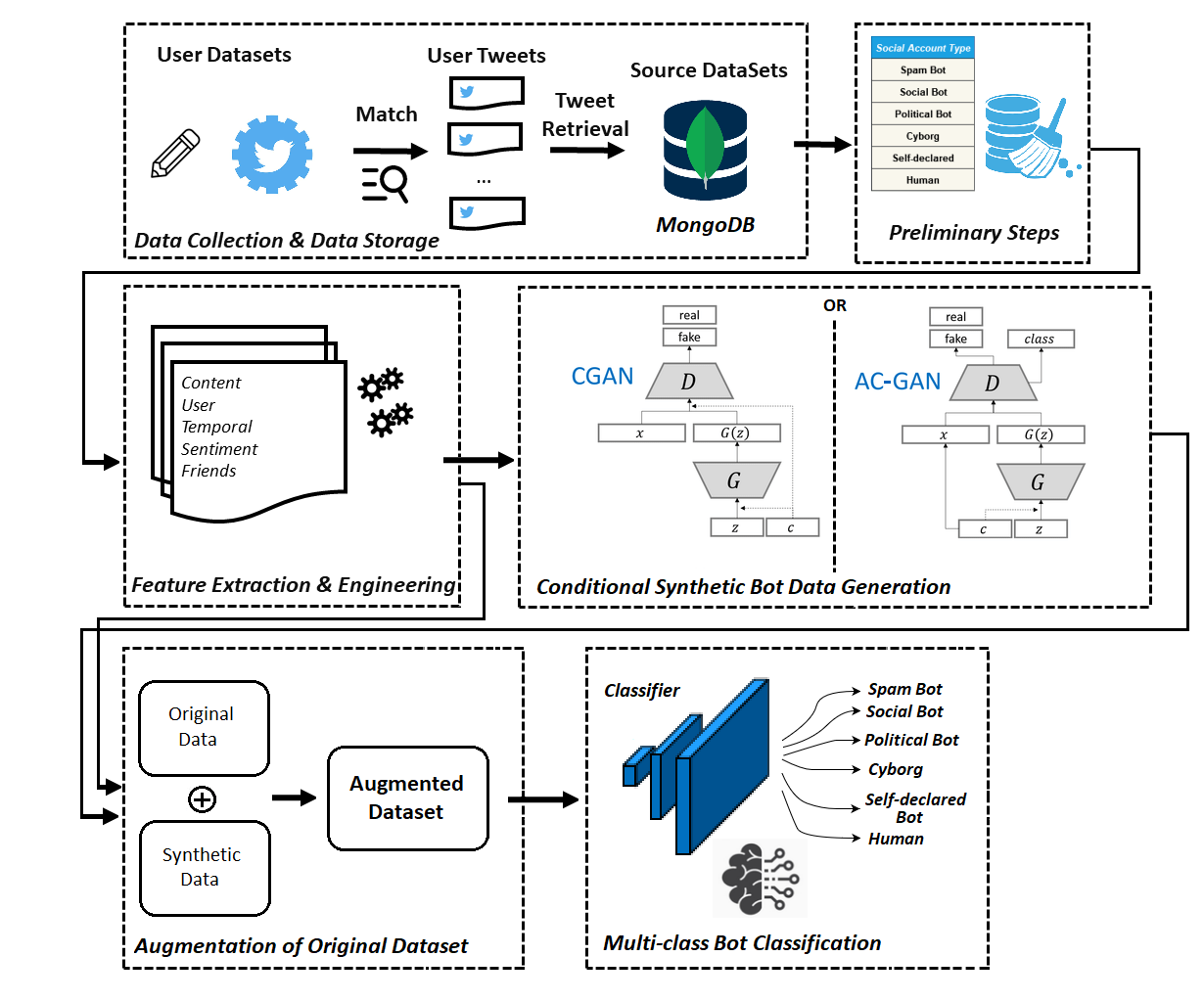}
    \caption{Architecture Pipeline of CALEB.}
    \label{fig:architecture}
\end{figure}

As outlined in Figure \ref{fig:architecture}, the proposed Bot Detection framework is composed of six distinct components: 
\begin{enumerate}
    \item \textbf{Data Collection \& Storage}: The first step of this work requires collecting the necessary data. As already mentioned in Section \ref{sec_dataset}, in this work we have used a dataset that was previously created in \cite{dimitriadis_botomics}, which consists of publicly available data and manually collected data from Twitter representing social accounts. 
    \item \textbf{Preliminary Steps}: The following step in our pipeline refers to the preprocessing of the data, as well as the multi-type bot categorization. Since in this work we are interested to address the issue of bot evolution in multi-type bot data, we have followed the bot categorization that is presented in Table \ref{bot_types_table}, as already discussed in Section \ref{sec_dataset}.
    \item \textbf{Feature Extraction \& Engineering}: The final step before training our deep learning adversarial models is to extract the necessary features. As described in Section \ref{sec_dataset}, each social account in our data is represented by 310 features that are divided in five categories: Temporal, Content, Sentiment, User, and Hashtag Correlation features.
    \item \textbf{Synthetic Bot Data Generation}: Having collected and pre-processed the required bot data, we train two GAN models which we describe in Section \ref{sec_gans}, namely Conditional Generative Adversarial Network (CGAN) and Auxiliary Classifier - Generative Adversarial Network (AC-GAN), to generate realistic synthetic bot instances of multiple types.
    \item \textbf{Data Augmentation}: The artificial bot data generated by GANs are then used to augment the original train and test sets, in a process we thoroughly describe in Section \ref{sec_bot_evol}.
    \item \textbf{Multi-type Bot Classification}: The last step in our pipeline requires training a ML classifier to perform the multi-type bot discrimination. For this purpose, we have decided to use a Random Forest classifier, as discussed in Section \ref{initial_setup}. In addition, instead of using an external ML classifier, we can use the Discriminator of AC-GAN, as we describe in Sections \ref{sec_ac_gan} and \ref{exp_discriminator}.
\end{enumerate}

Next, we thoroughly describe CGAN and AC-GAN, which we use to generate synthetic instances of bots. 

\subsection{Synthetic Bot Data Generation using GANs} \label{sec_gans}
Generative Adversarial Networks (GANs) have recently been introduced as a data augmentation technique, where a GAN is trained to generate realistic synthetic samples similar to original ones \cite{antoniou2017data}. These artificially produced data can then be utilized to augment the original dataset and help ML algorithms achieve better performance. However, in the original GAN, there is no control on modes, i.e., classes, of the data being generated. Therefore, \textbf{the idea behind using CGAN and AC-GAN to generate synthetic bot samples is that, in contrast with vanilla GAN, they offer the ability to create samples of specific classes}. In addition, the above GAN models provide a more stable training procedure than the simple GAN.

    \begin{figure}[t]
    \centering
    \includegraphics[scale=0.35]{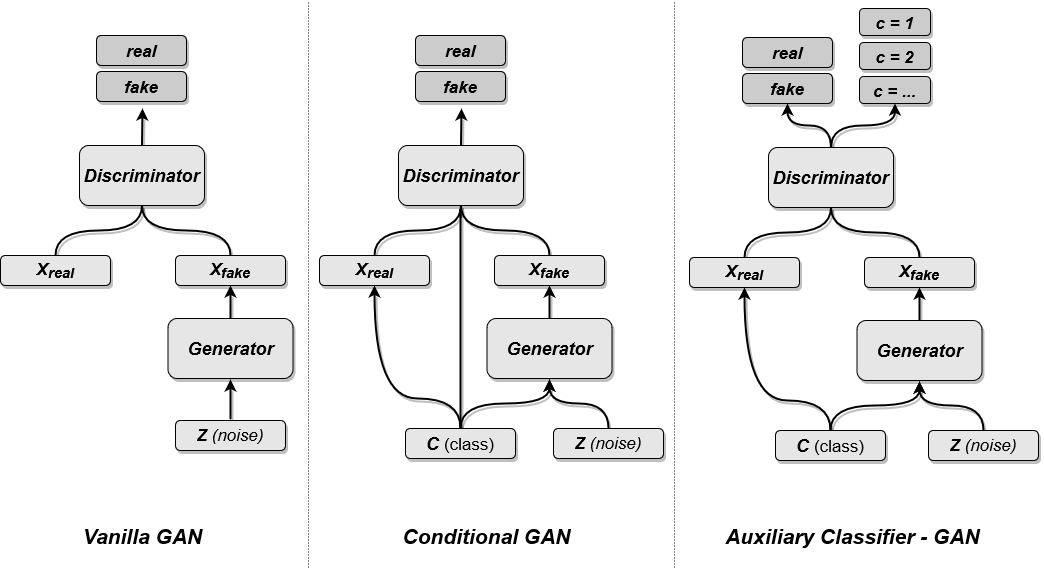}
    \caption{GAN Architectures Comparison.}
    \label{fig:gans}
    \end{figure}

\subsubsection{Conditional Generative Adversarial Networks}
Conditional GAN, introduced in \cite{mirza2014conditional}, is an extension of the vanilla GAN, as they both share almost the same model architecture as shown in Figure \ref{fig:gans}. However, the main difference of the CGAN is that it incorporates additional information regarding the class labels in the input data. Specifically, both the Generator (G) and Discriminator (D) are conditioned on auxiliary information $c$ by taking into account the class label for each sample in our training data. The information for the class labels is fed to both G and D as an additional Embedding input layer. In G, the Embedding layer with the class labels is concatenated with the prior input noise $p_{z}(z)$. In D, the Embedding layer with the class labels is concatenated with the data samples, either real or fake, and is given to the input layer.
Similar to the vanilla GAN, G and D are both trained simultaneously. The parameters of G are adjusted to minimize $\log (1-D(G(z \mid c))$, and the parameters of D are adjusted to minimize $\log (D(X \mid \boldsymbol{c}))$ as if they are following the two-player min-max game with value function $V (D, G)$:
\begin{equation}
\begin{split}   
\min _{G} \max _{D} V(D, G)=\mathbb{E}_{x \sim p_{\text {data }}(x)}[\log D(x \mid c)] + \mathbb{E}_{z \sim p_{x}(z)}[\log (1-D(G(z \mid c)))]
\end{split}
\end{equation}

\subsubsection{Auxiliary Classifier - Generative Adversarial Networks} \label{sec_ac_gan}
The Auxiliary Classifier - Generative Adversarial Network (AC-GAN) is an extension of the CGAN, introduced by A. Odena et al. \cite{odena2017conditional}. As presented in Figure \ref{fig:gans}, the main difference between CGAN and AC-GAN is in the discriminator model, which in the case of the AC-GAN, it is not fed fed with the class label of our data. Instead of receiving the class label as input, the discriminator is trained to predict it. Therefore, D has two outputs, one for the probability distribution over sources (training dataset or generator) and one for the probability distribution over the class labels. 


The objective function of the AC-GAN is composed of two parts: the log-likelihood of the correct source, $L_S$ , and the log-likelihood of the correct class, $L_C$.
\begin{equation}
L_{S}=E\left[\log P\left(S=\text { real } \mid X_{\text {real }}\right)\right] + E\left[\log P\left(S=\text { fake } \mid X_{\text {fake }}\right)\right]
\end{equation}

\begin{equation}
L_{C}=E\left[\log P\left(C=c \mid X_{\text {real }}\right)\right]+E\left[\log P\left(C=c \mid X_{\text {fake }}\right)\right]
\end{equation}

\noindent D is trained to maximize $L_S + L_C$ since its goal is to predict both the source of data and the class they belong to with the highest credibility possible. On the other hand, G is trained to maximize $L_C - L_S$, since its objective is to fool the discriminator into believing that generated data come from the training set while helping D map data to labels.


The most significant benefit of AC-GAN is that the discriminator can now be used as a classifier to detect the different types of social bots. Therefore, AC-GAN serves two purposes. First, we can use the generator of the AC-GAN to create \textbf{realistic synthetic bot instances}. Secondly, we can utilize the discriminator of the model for multi-class bot classification, \textbf{reducing the training overhead} that applies when training additional ML classifiers on the synthetic data for bot detection, in a process we illustrate in Section \ref{exp_discriminator}.

\section{Experimental Evaluation} \label{sec_experiments}
In this section, we describe the experimental procedure we followed, and we highlight the key observations and the main findings.

\subsection{Initial Setup} \label{initial_setup}
Following the state-of-the-art hyper-parameter tuning of CGAN and AC-GAN, our GAN models are composed of feed-forward neural networks for the Generator and Discriminator. We used a noise vector of size 128, and the size of the Embedding layer was set to 6 to match the number of classes in our data. The loss function for both networks was constructed using Binary Cross-Entropy loss. The two models were trained using stochastic gradient descent with the Adam optimizer and mini-batches of size 512 and a learning rate of 0.0002. We chose to train our models for 300 epochs, since the loss did not show to decrease any further after that point. The implementation was made using the PyTorch open-source machine learning library in Python. Finally, we decided to use a Random Forest (RF) classifier with default parameters to perform the classification, since it is the most widely used classifier in the bot detection literature. For all experiments we considered a 75\%-25\% train-test split.

\subsection{Handling Class Imbalance}
Before proceeding with the bot evolution results, we wanted to address the problem of class imbalance in our data, as shown in Table \ref{bot_types_table}. To this end, we used CGAN and AC-GAN to generate synthetic samples using the expansion multiple to augment the original training dataset. The expansion multiple is defined as: 
\begin{equation}
 \varphi=a: b  
\end{equation}
where $b$ is the number of social accounts per class in the training data, and $a$ is the number of samples generated by the CGAN or AC-GAN. Drawing intuition from \cite{wu2020using}, we have decided to apply $\varphi = 2:1$ in all classes, meaning that for each sample in the training set, we generate two synthetic examples. We trained RF on the CGAN augmented data and evaluated its performance on o hold-out test set. We compared CGAN augmented data to two other state-of-the-art oversampling techniques, namely ADASYN \cite{adasyn} and SMOTE-ENN \cite{smote}. In addition, we report the results when no imbalance handling technique is used, as presented in Table \ref{table_class_imbalance}. 

\begin{table}[h]
\centering
\caption{Class Imbalance Results using Random Forest with different augmentation techniques.}
\label{table_class_imbalance}
\resizebox{0.8\columnwidth}{!}{
\centering
\begin{tabular}{c|ccccc}
\begin{tabular}[c]{@{}c@{}}\textbf{Augmentation Technique}\end{tabular} & \textbf{Accuracy} & \textbf{Precision} & \textbf{F1 score} & \textbf{Recall} & \textbf{G-Mean}  \\ 
\hline
Original                                                                           & 0.8903            & 0.9057             & 0.8762            & 0.8520          & 0.9095           \\
ADASYN                                                                             & 0.8866            & 0.8570             & 0.8736            & 0.8922          & 0.9315           \\
SMOTE-ENN                                                                          & 0.8537            & 0.7940             & 0.8370            & 0.8996          & 0.9327           \\ 
\hline
CGAN $2:1$                                                                           & 0.8888            & 0.9069             & 0.8788            & 0.8518          & 0.9094          
\end{tabular}
}
\end{table}

\begin{remark}
\textbf{
Overall, we observe that ADASYN offers the best performance while our proposed CGAN method comes second best. In addition, we notice that Random Forest maintains a remarkably high performance even when no imbalance technique is used.}
\end{remark}
\noindent The above remark highlights that even though there is class imbalance in our data, no augmentation technique is necessary in order to maintain a high classification performance. 
\subsection{Synthetic Data Evaluation} \label{section_data_eval}

\begin{figure}[h]
    \centering
    \includegraphics[scale=0.55]{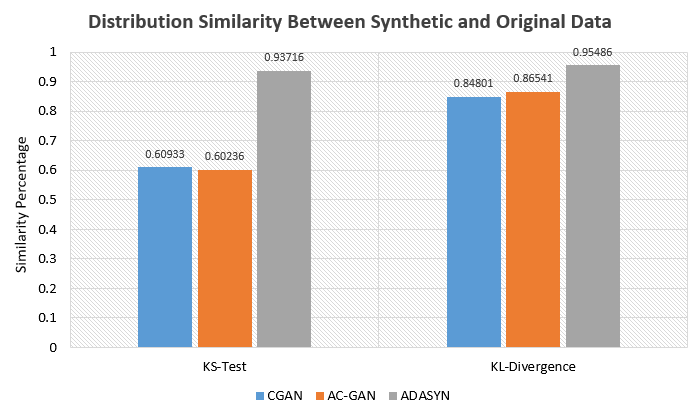}
    \caption{Distribution Similarity between Original and Synthetic Data of different Techniques.}
    \label{fig:synthetic_data_eval}
\end{figure}

\noindent Since the primary goal of this work is to tackle the challenge of bot evolution, we need to examine the quality of the synthetic data we are going to use for training and testing RF, and inspect the similarity between the original data and the artificial ones produced either by our Adversarial models (CGAN and AC-GAN) or ADASYN. For this reason, in Figure \ref{fig:synthetic_data_eval} we present the similarity between the original data and different sets of synthetic data (CGAN, AC-GAN, and ADASYN), using two synthetic data evaluation metrics from the Synthetic Data Vault \cite{sdv}, the Kolmogorov–Smirnov (KS) test and the continuous Kullback-Leibler (KL) Divergence metric. Both metrics calculate the difference between the distribution of the synthetic data and original data across all features and return the average which is a number between 0 and 1, with 1 meaning that the two distributions are identical.

\begin{remark}
\textbf{
We observe that ADASYN creates copies of the original data since its synthetic data have almost identical distribution to the original. On the other hand, CGAN and AC-GAN generate artificial data that feature some variety. In this way, the novel synthetic GAN data cover a broader range of bots that can simulate different types of evolving bots.}
\end{remark}

\noindent This remark points out the fact that not every augmentation technique can be used to simulate evolving bots, since it must create synthetic bot instances that introduce some variation compared to the original data. Our proposed approach achieves this variation between original and artificially made data emphasizing this work's \textbf{(}\ref{c2} contribution.

\subsection{Bot Evolution Results} \label{sec_bot_evol}
\begin{figure}[t]
    \centering
    \includegraphics[scale=0.5]{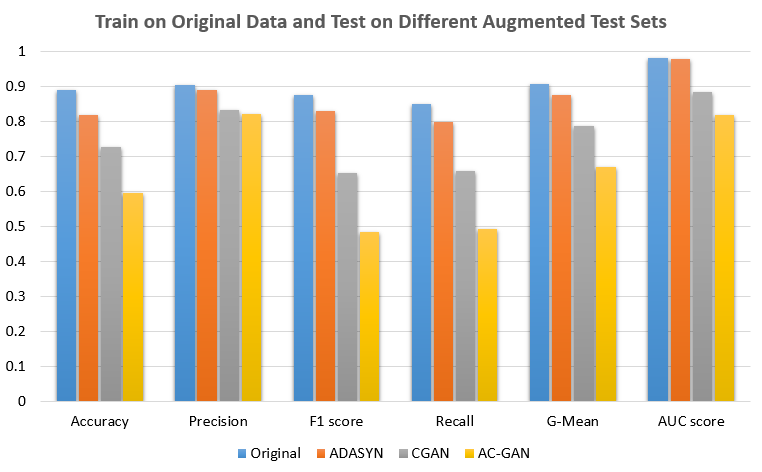}
    \caption{Evaluation on Different Augmented Test Sets by Training Random Forest only Original Data.}
    \label{fig:train_on_original}
\end{figure}

The first investigation we are interested in is the ability of a classifier trained only on existing bot data to detect future generations of bots. To this end, we simulate the evolution of bots by creating realistic synthetic bot instances using three different technique, two GAN models as described in Section \ref{sec_methodology}, and ADASYN. We then induce the synthetic data into the original test set, constructing different augmented test sets based on the technique that was used to generate the artificial data. Finally, we train a Random Forest classifier using only the original training data and we evaluate the model on the three augmented test sets, as described above. In addition, we report the results when we evaluate the model on the pure original test set. The results of this process are presented in Figure \ref{fig:train_on_original}.

\begin{figure}[b]
     \centering
     \subfigure[Precision]{\label{subfigure_precision}\includegraphics[scale=0.5]{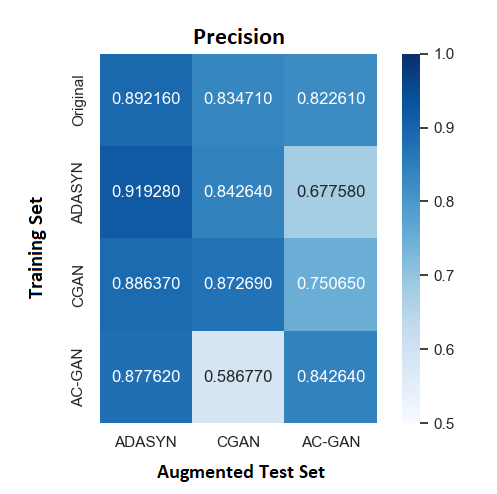}}
      \subfigure[G-mean]{\label{subfigure_g_mean}\includegraphics[scale=0.5]{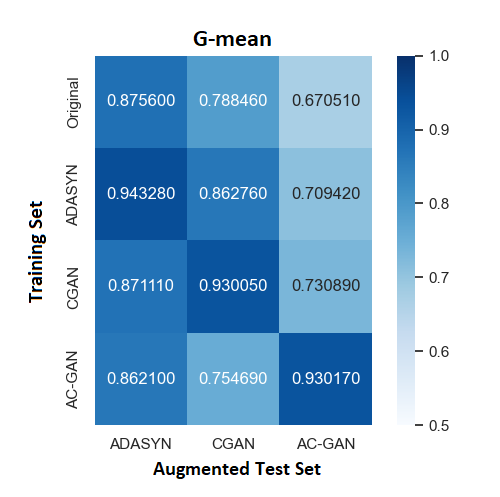}} 
      
    \caption{Random Forest's Precision and G-mean for different combinations of training and test sets. On the y-axis is the augmented training set and on the x-axis is the augmented test set.}
    \label{heatmaps}
\end{figure}

As can be observed, the classification performance of RF is greatly decreased when the original test set is augmented with synthetic data, i.e., when simulated bots make their appearance in the test set. ADASYN synthetic data do not diminish RF's performance by a big margin, which validates our observations as stated in Section \ref{section_data_eval}. Therefore, an ML classifier trained only on the original data can still classify ADASYN's data with acceptable performance, which is above 80\% in terms of all the evaluation metrics. On the other hand, \textbf{when we induce CGAN or AC-GAN synthetic data into the original test data, we observe a significant performance decrease across all evaluation metrics, with the biggest impact presented in F1-score and Recall which fall under 50\%}. This poor performance suggests that RF is not able to classify the simulated evolved bots and requires additional training information to be able to succeed.

Towards this end, we consider three different augmentation techniques for the training data, along with the pure original training set, as follows:
\begin{itemize}
    \item $ad_1$: original data augmented with \textbf{AC-GAN} synthetic data
    \item $ad_2$: original data augmented with \textbf{CGAN} synthetic data
    \item $ad_3$: original data augmented with \textbf{ADASYN} synthetic data
    \item $ad_4$: pure original data
\end{itemize}
The above training sets are illustrated on the $y$ axis in Figure \ref{heatmaps}.
It is worth mentioning that the 2:1 expansion multiple was used to generate synthetic data from CGAN and AC-GAN, since it proved to be the best technique during the class imbalance experimentation. In Figure \ref{heatmaps}, we present the Precision (\ref{subfigure_precision}) and G-mean (\ref{subfigure_g_mean}) that RF obtains under different combinations of training and test data. It is obvious that RF achieves the best performance when trained and tested on the same type of augmented data. However, RF underperforms especially when tested on CGAN and AC-GAN augmented data while it has been previously trained only on $ad_3$ or $ad_4$. On the contrary, when we train RF on $ad_1$ or $ad_2$ and test on ADASYN augmented data, RF achieves relatively high performance, as shown in Figure \ref{heatmaps}.

So far, the experimental evaluation on different combinations of augmented train and test sets has led to the following observations: 
\begin{remark}
\textbf{ADASYN's synthetic data are not useful for simulating evolving social bots, since they include copies of the original data and can be easily classified by RF trained only on the original data.}
\end{remark}

\begin{remark}
\textbf{CGAN and AC-GAN synthetic data greatly decrease RF's performance when it is trained on other training sets, with AC-GAN synthetic data being the most difficult to classify correctly.}
\end{remark}

\noindent
Up to now, to validate the efficiency of our proposed approach we have used simulated evolved bots. However, this is not the ideal way to evaluate our method, since we are not fully convinced whether our GAN-based methodology accurately simulates bot evolution. To this end, we have decided to proceed with further experimentation based on a real world scenario, as described below.
\\ \\
\noindent \textbf{Real Evolved Bot Data Experimentation:}\\ \\
To further verify that our proposed adversarial methodology can effectively simulate bot evolution and be used proactively to detect future generations of bots, we considered a real-world example using binary bot data. Initially, we trained a Random Forest classifier on an old public dataset (Caverlee 2011 \cite{lee2011seven_4}). We then evaluated this model on three more recent datasets (Gilani 2017 \cite{diesner2017proceedings_62}, Varol 2017 \cite{varol2017online}, and Cresci\_Stock 2018 \cite{cresci2018fake_18}), that include evolved bot instances. We proceeded by augmenting the original data with AC-GAN using the 2:1 expansion multiple as in the above experiments. The idea behind this experiment is that the more recent datasets contain newer evolved bots that do not exist in the older dataset. 

As can be seen in Table \ref{binary_table}, augmenting the original training data with AC-GAN always boosts the performance of Random Forest, regardless of which one of the three newer datasets is our test set. At this point, we should mention that the absolute scores are not so crucial since the test datasets contain different types of bots, and we only focus on binary classification. On the contrary, we are interested in the performance boost that our proposed methodology offers. For instance, \textbf{we obtain a performance boost of almost 10\% when we evaluate RF on the Cresci\_Stock \cite{cresci2018fake_18} dataset which is the most recent among the ones we have considered for this experiment}. This reveals that our adversarial methodology can effectively generate realistic synthetic bot samples that simulate evolving bots and help towards the detection of their future generations before they even emerge, once again highlighting this paper's \textbf{(}\ref{c2} contribution. In addition, in Table \ref{binary_table}, we include the results when we augment the original training data with ADASYN. 

\begin{table}[h]
\centering
\caption{Random Forest performance trained on an older bot dataset (\cite{lee2011seven_4} 2011) and its augmented variations and evaluated on newer datasets (\cite{diesner2017proceedings_62} 2017, \cite{varol2017online} 2017, \cite{cresci2018fake_18} 2018).}
\label{binary_table}
\begin{tabular}{c|c|ccccc}
\textbf{Test Data}                              & \textbf{Train Data}                                       & \textbf{Accuracy}                                    & \textbf{Precision}                                   & \textbf{F1 score}                                    & \textbf{Recall}                                      & \textbf{G-Mean}                                       \\ 
\hline
\multirow{3}{*}{\textbf{Gilani (2017) \cite{diesner2017proceedings_62}}}        & {\cellcolor[rgb]{0.839,0.839,0.839}}Caverlee (2011) \cite{lee2011seven_4}       & {\cellcolor[rgb]{0.839,0.839,0.839}}0.59298          & {\cellcolor[rgb]{0.839,0.839,0.839}}0.58490          & {\cellcolor[rgb]{0.839,0.839,0.839}}0.53621          & {\cellcolor[rgb]{0.839,0.839,0.839}}0.55783          & {\cellcolor[rgb]{0.839,0.839,0.839}}0.48260           \\
                                                & Augmented with ADASYN                                      & 0.58802                                              & 0.58356                                              & 0.51006                                              & 0.54654                                              & 0.43546                                               \\
                                                & {\cellcolor[rgb]{0.839,0.839,0.839}}Augmented with AC-GAN & {\cellcolor[rgb]{0.839,0.839,0.839}}\textbf{0.61279} & {\cellcolor[rgb]{0.839,0.839,0.839}}\textbf{0.61620} & {\cellcolor[rgb]{0.839,0.839,0.839}}\textbf{0.55588} & {\cellcolor[rgb]{0.839,0.839,0.839}}\textbf{0.57693} & {\cellcolor[rgb]{0.839,0.839,0.839}}\textbf{0.50251}  \\ 
\hline
\multirow{3}{*}{\textbf{Varol (2017) \cite{varol2017online}}}         & Caverlee (2011) \cite{lee2011seven_4}                                           & 0.79186                                              & 0.76903                                              & 0.76105                                              & 0.75521                                              & 0.74686                                               \\
                                                & {\cellcolor[rgb]{0.839,0.839,0.839}}Augmented with ADASYN  & {\cellcolor[rgb]{0.839,0.839,0.839}}0.78465          & {\cellcolor[rgb]{0.839,0.839,0.839}}0.76638          & {\cellcolor[rgb]{0.839,0.839,0.839}}0.74430          & {\cellcolor[rgb]{0.839,0.839,0.839}}0.73317          & {\cellcolor[rgb]{0.839,0.839,0.839}}0.71608           \\
                                                & Augmented with AC-GAN                                     & \textbf{0.80732}                                     & \textbf{0.78533}                                     & \textbf{0.78169}                                     & \textbf{0.77856}                                     & \textbf{0.76927}                                      \\ 
\hline
\multirow{3}{*}{\begin{tabular}[c]{@{}c@{}}\textbf{Cresci\_Stock }\\\textbf{(2018) \cite{cresci2018fake_18} }\end{tabular}}  & {\cellcolor[rgb]{0.839,0.839,0.839}}Caverlee (2011) \cite{lee2011seven_4}       & {\cellcolor[rgb]{0.839,0.839,0.839}}0.57139          & {\cellcolor[rgb]{0.839,0.839,0.839}}0.58350          & {\cellcolor[rgb]{0.839,0.839,0.839}}0.56933          & {\cellcolor[rgb]{0.839,0.839,0.839}}0.58011          & {\cellcolor[rgb]{0.839,0.839,0.839}}0.56885           \\
                                                & Augmented with ADASYN                                      & 0.59539                                              & 0.60676                                              & 0.59402                                              & 0.60339                                              & 0.59429                                               \\
                                                & {\cellcolor[rgb]{0.839,0.839,0.839}}Augmented with AC-GAN & {\cellcolor[rgb]{0.839,0.839,0.839}}\textbf{0.62840} & {\cellcolor[rgb]{0.839,0.839,0.839}}\textbf{0.63545} & {\cellcolor[rgb]{0.839,0.839,0.839}}\textbf{0.62816} & {\cellcolor[rgb]{0.839,0.839,0.839}}\textbf{0.63405} & {\cellcolor[rgb]{0.839,0.839,0.839}}\textbf{0.61708} 
\end{tabular}
\end{table}

The results show that not only ADASYN does not offer any improvement in the performance but in many cases it degrades it, confirming that ADASYN is not a suitable method for simulating evolving bots and showcasing that not all techniques that generate synthetic data can simulate bot evolution.

\subsection{Evaluating AC-GAN as a Bot Detector} \label{exp_discriminator}
So far, throughout our experimental evaluation, we have considered only Random Forest as our bot detector to perform the classification of the different types of bots. However, this approach requires to first construct the augmented training datasets and then train additional ML classifiers, such as RF, to perform the classification, adding extra overhead in our pipeline. An alternative approach is to directly use the already trained Discriminator network of the AC-GAN for multi-class classification. 

\begin{figure}[h]
    \centering
    \includegraphics[scale=0.55]{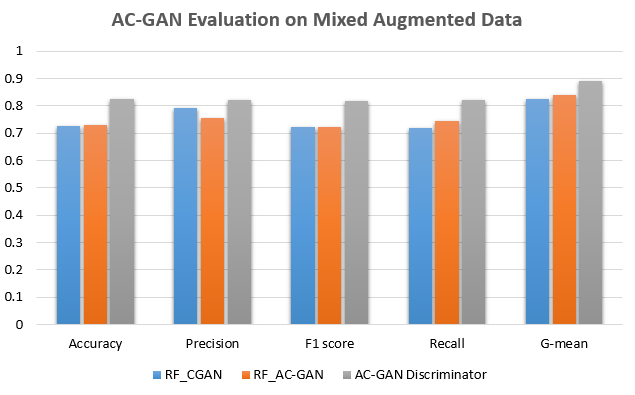}
    \caption{Performance Comparison between Random Forest (RF) and Discriminator of AC-GAN on Mixed Augmented Data.}
    \label{fig:ac_gan_bot_detector}
\end{figure}

In Figure \ref{fig:ac_gan_bot_detector}, we present a performance comparison between RF and AC-GAN's Discriminator on mixed augmented data, which consist of original data, CGAN and AC-GAN synthetic data. The idea behind using a mixed augmented dataset is that we want to provide a fair evaluation of the classification models, since evaluating the discriminator of AC-GAN only on AC-GAN augmented data would obviously provide better results than RF. In this experiment, RF is trained once with CGAN augmented data, denoted by RF\_CGAN, and once with AC-GAN augmented data, denoted by RF\_AC-GAN. On the other hand, AC-GAN only uses the original training data to train both the Generator and the Discriminator. As can be observed, AC-GAN's Discriminator outperforms both RF with CGAN and AC-GAN data.

\begin{remark}
\textbf{
The already trained Discriminator of AC-GAN can be used for the classification of multi-type evolved bots, omitting the need for additional ML classifiers. Therefore, AC-GAN composes an end-to-end bot detection framework which is robust against evolving bots, serving two purposes, generating synthetic bot data and successfully classifying existing and evolving social bots.}
\end{remark}
\noindent The above results and observation justify the most important contribution of this paper \textbf{(}\ref{c3}, since to the best of our knowledge, this is the first time a GAN is used for classification in the bot detection domain.
\section{Conclusion and Future Work} \label{sec_conclusion}
In this paper, we propose CALEB, a Conditional Adversarial Learning Framework to proactively detect multi-type evolving bots in Online Social Networks. In this context, we employed two GAN models, namely CGAN and AC-GAN, which were trained to create realistic synthetic bot instances of multiple types. The artificial GAN data represented evolved versions of bots and were used to augment the existing bot datasets in order to construct a robust ML classifier against future generations of bots. 

Results showed that CALEB can effectively simulate evolving bots and help ML models detect future generations of bots with better performance compared to previous works. Moreover, our experimental analysis showed that other augmentation techniques that are widely used in class imbalance problems, such as ADASYN, are not suitable for simulating evolving bots, since they create synthetic data that are almost identical to the original one. Finally, we evaluated the Discriminator of AC-GAN as a bot detector, which showed to outperform Random Forest, revealing that there is no need to train additional classifiers to perform the multi-type detection of evolving bots.

Future work may focus on creating synthetic bot samples that present specific modified features using Controllable GANs \cite{controllable}, by leveraging the latent space disentanglement properties of the GAN. In this way, we may be able to simulate evolved bots in a more accurate way. In addition, an interesting idea for the future is to construct a robust set of language-agnostic features to overcome the limitation that is presented in most existing public bot datasets by non-English Twitter content.

\bibliographystyle{unsrt}  
\bibliography{references}

\begin{thebibliography}{10}

\bibitem{kepios_general}
Global social media stats, retrieved from:
  https://datareportal.com/social-media-users, 2021.

\bibitem{kepios_twitter}
Essential twitter stats for 2021, retrieved from:
  https://datareportal.com/essential-twitter-stats, 2021.

\bibitem{ferrara2016rise}
Emilio Ferrara, Onur Varol, Clayton Davis, Filippo Menczer, and Alessandro
  Flammini.
\newblock The rise of social bots.
\newblock {\em Communications of the ACM}, 59(7):96--104, 2016.

\bibitem{botomics2}
Stefan Stieglitz, Florian Brachten, Bj{\"o}rn Ross, and Anna-Katharina Jung.
\newblock Do social bots dream of electric sheep? a categorisation of social
  media bot accounts.
\newblock {\em arXiv preprint arXiv:1710.04044}, 2017.

\bibitem{botomics21}
Victor Luckerson.
\newblock Can twitter solve its big, bad bot problem?. available at:
  https://www.theringer.com/tech/2018/3/8/1 7093982/twitter-bot-problem, 2018.

\bibitem{cgan5}
Christopher~A Cassa, Rumi Chunara, Kenneth Mandl, and John~S Brownstein.
\newblock Twitter as a sentinel in emergency situations: lessons from the
  boston marathon explosions.
\newblock {\em PLoS currents}, 5, 2013.

\bibitem{wu2020using}
Bin Wu, Le~Liu, Yanqing Yang, Kangfeng Zheng, and Xiujuan Wang.
\newblock Using improved conditional generative adversarial networks to detect
  social bots on twitter.
\newblock {\em IEEE Access}, 8:36664--36680, 2020.

\bibitem{cgan8}
Norah Abokhodair, Daisy Yoo, and David~W McDonald.
\newblock Dissecting a social botnet: Growth, content and influence in twitter.
\newblock In {\em Proceedings of the 18th ACM conference on computer supported
  cooperative work \& social computing}, pages 839--851, 2015.

\bibitem{cmucovid}
Virginia~Alvino Young.
\newblock Nearly half of the twitter accounts discussing 'reopening america'
  may be bots, 2020.

\bibitem{ferrara2020covid}
Emilio Ferrara.
\newblock \# covid-19 on twitter: Bots, conspiracies, and social media
  activism.
\newblock {\em arXiv preprint arXiv: 2004.09531}, 2020.

\bibitem{botomics14}
David~A Broniatowski, Amelia~M Jamison, SiHua Qi, Lulwah AlKulaib, Tao Chen,
  Adrian Benton, Sandra~C Quinn, and Mark Dredze.
\newblock Weaponized health communication: Twitter bots and russian trolls
  amplify the vaccine debate.
\newblock {\em American journal of public health}, 108(10):1378--1384, 2018.

\bibitem{botomics15}
Despoina Chatzakou, Nicolas Kourtellis, Jeremy Blackburn, Emiliano
  De~Cristofaro, Gianluca Stringhini, and Athena Vakali.
\newblock Mean birds: Detecting aggression and bullying on twitter.
\newblock In {\em Proceedings of the 2017 ACM on web science conference}, pages
  13--22, 2017.

\bibitem{cgan9}
Venkatramanan~S Subrahmanian, Amos Azaria, Skylar Durst, Vadim Kagan, Aram
  Galstyan, Kristina Lerman, Linhong Zhu, Emilio Ferrara, Alessandro Flammini,
  and Filippo Menczer.
\newblock The darpa twitter bot challenge.
\newblock {\em Computer}, 49(6):38--46, 2016.

\bibitem{cresci2020decade}
Stefano Cresci.
\newblock A decade of social bot detection.
\newblock {\em Communications of the ACM}, 63(10):72--83, 2020.

\bibitem{cresci2019better}
Stefano Cresci, Marinella Petrocchi, Angelo Spognardi, and Stefano Tognazzi.
\newblock Better safe than sorry: An adversarial approach to improve social bot
  detection.
\newblock In {\em Proceedings of the 10th ACM Conference on Web Science}, pages
  47--56, 2019.

\bibitem{decade5}
Stefano Cresci, Roberto Di~Pietro, Marinella Petrocchi, Angelo Spognardi, and
  Maurizio Tesconi.
\newblock The paradigm-shift of social spambots: Evidence, theories, and tools
  for the arms race.
\newblock In {\em Proceedings of the 26th international conference on world
  wide web companion}, pages 963--972, 2017.

\bibitem{bettersafe43}
Stefano Cresci, Marinella Petrocchi, Angelo Spognardi, and Stefano Tognazzi.
\newblock From reaction to proaction: Unexplored ways to the detection of
  evolving spambots.
\newblock In {\em Companion Proceedings of the The Web Conference 2018}, pages
  1469--1470, 2018.

\bibitem{cresci2021coming}
Stefano Cresci, Marinella Petrocchi, Angelo Spognardi, and Stefano Tognazzi.
\newblock The coming age of adversarial social bot detection.
\newblock {\em First Monday}, 2021.

\bibitem{goodfellow2014generative}
Ian Goodfellow, Jean Pouget-Abadie, Mehdi Mirza, Bing Xu, David Warde-Farley,
  Sherjil Ozair, Aaron Courville, and Yoshua Bengio.
\newblock Generative adversarial nets.
\newblock In {\em Advances in neural information processing systems}, pages
  2672--2680, 2014.

\bibitem{cresci2016dna}
Stefano Cresci, Roberto Di~Pietro, Marinella Petrocchi, Angelo Spognardi, and
  Maurizio Tesconi.
\newblock Dna-inspired online behavioral modeling and its application to
  spambot detection.
\newblock {\em IEEE Intelligent Systems}, 31(5):58--64, 2016.

\bibitem{dimitriadis_botomics}
Ilias Dimitriadis, Konstantinos Georgiou, and Athena Vakali.
\newblock Social botomics: A systematic ensemble ml approach for explainable
  and multi-class bot detection.
\newblock {\em Applied Sciences}, 11(21):9857, 2021.

\bibitem{davis2016botornot}
Clayton~Allen Davis, Onur Varol, Emilio Ferrara, Alessandro Flammini, and
  Filippo Menczer.
\newblock Botornot: A system to evaluate social bots.
\newblock In {\em Proceedings of the 25th international conference companion on
  world wide web}, pages 273--274, 2016.

\bibitem{lee2010uncovering}
Kyumin Lee, James Caverlee, and Steve Webb.
\newblock Uncovering social spammers: social honeypots+ machine learning.
\newblock In {\em Proceedings of the 33rd international ACM SIGIR conference on
  Research and development in information retrieval}, pages 435--442, 2010.

\bibitem{varol2017online}
Onur Varol, Emilio Ferrara, Clayton Davis, Filippo Menczer, and Alessandro
  Flammini.
\newblock Online human-bot interactions: Detection, estimation, and
  characterization.
\newblock In {\em Proceedings of the international AAAI conference on web and
  social media}, volume~11, 2017.

\bibitem{yang2019arming}
Kai-Cheng Yang, Onur Varol, Clayton~A Davis, Emilio Ferrara, Alessandro
  Flammini, and Filippo Menczer.
\newblock Arming the public with artificial intelligence to counter social
  bots.
\newblock {\em Human Behavior and Emerging Technologies}, 1(1):48--61, 2019.

\bibitem{yang2020scalable_8}
Kai-Cheng Yang, Onur Varol, Pik-Mai Hui, and Filippo Menczer.
\newblock Scalable and generalizable social bot detection through data
  selection.
\newblock In {\em Proceedings of the AAAI Conference on Artificial
  Intelligence}, volume~34, pages 1096--1103, 2020.

\bibitem{yardi2010detecting}
Sarita Yardi, Daniel Romero, Grant Schoenebeck, et~al.
\newblock Detecting spam in a twitter network.
\newblock {\em First monday}, 2010.

\bibitem{chu2012detecting}
Zi~Chu, Steven Gianvecchio, Haining Wang, and Sushil Jajodia.
\newblock Detecting automation of twitter accounts: Are you a human, bot, or
  cyborg?
\newblock {\em IEEE Transactions on dependable and secure computing},
  9(6):811--824, 2012.

\bibitem{kurakin2016adversarial}
Alexey Kurakin, Ian Goodfellow, and Samy Bengio.
\newblock Adversarial machine learning at scale.
\newblock {\em arXiv preprint arXiv:1611.01236}, 2016.

\bibitem{wu2019detecting}
Bin Wu, Le~Liu, Zhengge Dai, Xiujuan Wang, and Kangfeng Zheng.
\newblock Detecting malicious social robots with generative adversarial
  networks.
\newblock {\em KSII Transactions on Internet and Information Systems (TIIS)},
  13(11):5594--5615, 2019.

\bibitem{mirza2014conditional}
Mehdi Mirza and Simon Osindero.
\newblock Conditional generative adversarial nets.
\newblock {\em arXiv preprint arXiv:1411.1784}, 2014.

\bibitem{cresci2017social}
Stefano Cresci, Roberto Di~Pietro, Marinella Petrocchi, Angelo Spognardi, and
  Maurizio Tesconi.
\newblock Social fingerprinting: detection of spambot groups through
  dna-inspired behavioral modeling.
\newblock {\em IEEE Transactions on Dependable and Secure Computing},
  15(4):561--576, 2017.

\bibitem{miller2014twitter}
Zachary Miller, Brian Dickinson, William Deitrick, Wei Hu, and Alex~Hai Wang.
\newblock Twitter spammer detection using data stream clustering.
\newblock {\em Information Sciences}, 260:64--73, 2014.

\bibitem{jan2020throwing}
Steve~TK Jan, Qingying Hao, Tianrui Hu, Jiameng Pu, Sonal Oswal, Gang Wang, and
  Bimal Viswanath.
\newblock Throwing darts in the dark? detecting bots with limited data using
  neural data augmentation.
\newblock In {\em 2020 IEEE Symposium on Security and Privacy (SP)}, pages
  1190--1206. IEEE, 2020.

\bibitem{yin2018enhancing}
Chuanlong Yin, Yuefei Zhu, Shengli Liu, Jinlong Fei, and Hetong Zhang.
\newblock An enhancing framework for botnet detection using generative
  adversarial networks.
\newblock In {\em 2018 International Conference on Artificial Intelligence and
  Big Data (ICAIBD)}, pages 228--234. IEEE, 2018.

\bibitem{ma2019detect}
Jing Ma, Wei Gao, and Kam-Fai Wong.
\newblock Detect rumors on twitter by promoting information campaigns with
  generative adversarial learning.
\newblock In {\em The World Wide Web Conference}, pages 3049--3055, 2019.

\bibitem{yang2019arming_5}
Kai-Cheng Yang, Onur Varol, Clayton~A Davis, Emilio Ferrara, Alessandro
  Flammini, and Filippo Menczer.
\newblock Arming the public with artificial intelligence to counter social
  bots.
\newblock {\em Human Behavior and Emerging Technologies}, 1(1):48--61, 2019.

\bibitem{davis2016botornot_42}
Clayton~Allen Davis, Onur Varol, Emilio Ferrara, Alessandro Flammini, and
  Filippo Menczer.
\newblock Botornot: A system to evaluate social bots.
\newblock In {\em Proceedings of the 25th international conference companion on
  world wide web}, pages 273--274, 2016.

\bibitem{pca}
Karl~Pearson F.R.S.
\newblock Liii. on lines and planes of closest fit to systems of points in
  space.
\newblock {\em The London, Edinburgh, and Dublin Philosophical Magazine and
  Journal of Science}, 2(11):559--572, 1901.

\bibitem{antoniou2017data}
Antreas Antoniou, Amos Storkey, and Harrison Edwards.
\newblock Data augmentation generative adversarial networks.
\newblock {\em arXiv preprint arXiv:1711.04340}, 2017.

\bibitem{odena2017conditional}
Augustus Odena, Christopher Olah, and Jonathon Shlens.
\newblock Conditional image synthesis with auxiliary classifier gans.
\newblock In {\em International conference on machine learning}, pages
  2642--2651. PMLR, 2017.

\bibitem{adasyn}
Haibo He, Yang Bai, Edwardo~A Garcia, and Shutao Li.
\newblock Adasyn: Adaptive synthetic sampling approach for imbalanced learning.
\newblock In {\em 2008 IEEE international joint conference on neural networks
  (IEEE world congress on computational intelligence)}, pages 1322--1328. IEEE,
  2008.

\bibitem{smote}
Gustavo~EAPA Batista, Ronaldo~C Prati, and Maria~Carolina Monard.
\newblock A study of the behavior of several methods for balancing machine
  learning training data.
\newblock {\em ACM SIGKDD explorations newsletter}, 6(1):20--29, 2004.

\bibitem{sdv}
N.~{Patki}, R.~{Wedge}, and K.~{Veeramachaneni}.
\newblock The synthetic data vault.
\newblock In {\em 2016 IEEE International Conference on Data Science and
  Advanced Analytics (DSAA)}, pages 399--410, Oct 2016.

\bibitem{lee2011seven_4}
Kyumin Lee, Brian~David Eoff, and James Caverlee.
\newblock Seven months with the devils: A long-term study of content polluters
  on twitter.
\newblock In {\em Fifth international AAAI conference on weblogs and social
  media}, 2011.

\bibitem{diesner2017proceedings_62}
Jana Diesner, Elena Ferrari, and Guandong Xu.
\newblock {\em Proceedings of the 2017 IEEE/ACM International Conference on
  Advances in Social Networks Analysis and Mining 2017}.
\newblock 2017.

\bibitem{cresci2018fake_18}
Stefano Cresci, Fabrizio Lillo, Daniele Regoli, Serena Tardelli, and Maurizio
  Tesconi.
\newblock Fake: Evidence of spam and bot activity in stock microblogs on
  twitter.
\newblock In {\em Twelfth international AAAI conference on web and social
  media}, 2018.

\bibitem{controllable}
Alon Shoshan, Nadav Bhonker, Igor Kviatkovsky, and Gerard Medioni.
\newblock Gan-control: Explicitly controllable gans.
\newblock In {\em Proceedings of the IEEE/CVF International Conference on
  Computer Vision}, pages 14083--14093, 2021.

\end{thebibliography}

\end{document}